\documentclass[twocolumn,prd,superscriptaddress,preprintnumbers,nofootinbib]{revtex4}
\usepackage{graphicx,epsfig}
\usepackage{bm}
\usepackage{latexsym,amssymb,amsmath,float}
\usepackage{color}
\def\xe{{\epsilon}}

\def\deg{\ifmmode{^{\circ}}\else ${^{\circ}}$\fi}
\def\bi{\begin{itemize}}
\def\ei{\end{itemize}}
\def\bfl{\begin{flushleft}}
\def\efl{\end{flushleft}}
\def\ed{\end{document}}

\def\cf#1{\ifmmode{\cal #1}\else${\cal #1}$\fi}
\def\ra{\rightarrow}
\def\be{\begin{equation}}
\def\ee{\end{equation}}

\def\beas{\begin{eqnarray*}}
\def\eeas{\end{eqnarray*}}
\def\bea{\begin{eqnarray}}
\def\eea{\end{eqnarray}}

\def\mpl{M_{\rm Pl}}
\def\bm{\boldmath}
\newcommand{\zt}[1]{\textrm{#1}}

\newcommand{\postscript}[2]{\setlength{\epsfxsize}{#2\hsize}
   \centerline{\epsfbox{#1}}}

\newcommand{\er}[1]{\eqref{#1}}

\newcommand{\lb}{\left(}
\newcommand{\rb}{\right)}

\newcommand{\lsb}{\left[}
\newcommand{\rsb}{\right]}

\newcommand{\nn}{\nonumber \\}

\definecolor{rossoCP3}{cmyk}{0,.88,.77,.40}
\begin{document}

\title{\color{rossoCP3}{\bm${S}$-dual Inflation: BICEP2 data without unlikeliness}} 

\author{Luis A.~Anchordoqui}

\affiliation{Department of Physics and Astronomy, Lehman College at
  CUNY, 
Bronx NY 10468, USA}

\affiliation{Department of Physics,
University of Wisconsin-Milwaukee,
 Milwaukee, WI 53201, USA
}

\author{Vernon Barger}
\affiliation{Department of Physics, University of Wisconsin, Madison, WI 53706, USA}

\author{Haim Goldberg}
\affiliation{Department of Physics,
Northeastern University, Boston, MA 02115, USA
}

\author{Xing Huang}
\affiliation{Department of Physics, 
National Taiwan Normal University, Taipei, 116, Taiwan}

\author{Danny Marfatia}
\affiliation{Department of Physics and Astronomy,
University of Hawaii, Honolulu, HI 96822, USA}

\begin{abstract}
\noindent We show that $S$-dual inflationary potentials solve the unlikeliness problem manifested in Planck data and explain the excess $B$-mode power 
observed by the BICEP2 experiment as arising from primordial tensor fluctuations.
\end{abstract}
\maketitle

The discovery of primordial gravitational waves in the $B$-mode power
spectrum is fertile testing ground for inflationary
models~\cite{Ade:2014xna}. The $B$-mode power spectrum recently
reported by the BICEP2 collaboration is well-fit by a
lensed-$\Lambda$CDM + tensor model, with a tensor-to-scalar
ratio $r = 0.20^{+0.07}_{-0.05}$ and is inconsistent with the null
hypothesis, $r=0$, at a significance of $6-7\sigma$ depending on the modeling of foreground dust. Furthermore, the
new BICEP2 data when combined with observations from the Wilkinson
Microwave Anisotropy Probe (WMAP)~\cite{Hinshaw:2012aka}, the Atacama Cosmology Telescope
(ACT)~\cite{Sievers:2013ica}, the South Pole Telescope (STP)~\cite{Keisler:2011aw}, and the Planck
mission~\cite{Ade:2013uln} significantly shrink the space of allowed
inflationary cosmologies. 

Of particular interest here is that Planck data favor standard slow-roll
single field inflationary models with  plateau-like potentials $V(\phi)$ for which
$V'' <0$, over power-law potentials.
However, most of these plateau-like inflaton potentials
experience the so-called ``unlikeliness
problem''~\cite{Ijjas:2013vea}.  As an illustration of this problem
consider the simplest example of a plateau-like potential,
\begin{equation}
V(\phi) = \frac{V_0}{(2M)^4} \, [\phi^2- (2M)^2]^2 \, , 
\label{V1}
\end{equation}
where $V_0$ and $M$ are free parameters.  It is seen by inspection
that the plateau terminates at a local minimum, and then for large
$\phi$, the potential grows as a power-law $\sim V_0 \phi^4/
(2M)^4$. A problem arises because the minimum of the potential can
be reached in two different ways: by slow-roll along the
plateau or by slow-roll from the power-law side of the
minimum.  The path from the power-law side
requires less fine tuning of parameters, has inflation occuring over a much wider
range of $\phi$, and produces exponentially more inflation. Yet Planck data prefer
the unlikely path along the plateau. 
 
Currently, the combination of BICEP2 and Planck data does not disfavor
$V''>0$.  If in fact additional foregrounds are
present~\cite{Liu:2014mpa} that contribute to the $B$-mode power
observed by BICEP2, then potentials with $V''>0$ will be in tension
with the combination of these datasets.  We entertain this
possibility. From the perspective of existing data, we are interested
in solving the unlikeliness problem for those potentials with $V''<0$.

The requirement that $V'' < 0$ in the de Sitter region, and the
avoidance of the unlikeliness problem, must now also accommodate the
tensor-to-scalar ratio detected by BICEP2 data. Also, we would
like the inflaton potential to possess
some connection to particle physics. To this end, we hypothesize that
the potential be invariant under the $S$-duality constraint $g\ra 1/g$, or
$\phi\ra -\phi$, where $\phi$ is the dilaton/inflaton, and $g\sim
e^{\phi/M}.$ Here $M$ is expected to be within a few orders of
magnitude of the Planck mass $M_{\rm Pl}= G^{-1/2}$. (Throughout we use
  natural units, $c = \hbar \equiv 1$.)  This requirement forces the functional
form $V(\phi) = f[ \cosh(\phi/M)] $ on the potential. In what follows we
take for $V$ the simplest $S$ self-dual form 
\begin{equation}
V=V_0\ {\rm sech}(\phi/M) \,,
\label{V2}
\end{equation}
which solves the unlikeliness problem because it has no power-law wall.

$S$ duality had its origins in the Dirac quantization condition
on the electric and magnetic charges, suggesting an equivalence in the
description of quantum electrodynamics as either a weakly coupled theory of electric
charges or a strongly coupled theory of magnetic monopoles. This was
developed in~\cite{Montonen:1977sn} and extended into the
$S$-dualities of Type IIB string theories~\cite{Font:1990gx}.
We don't attempt a full association with a particular string
vacuum, but simply regard the self-dual constraint as a relic of
string physics in big bang cosmology.

For the potential in (\ref{V2}), the usual slow-roll
parameters are~\cite{Lidsey:1995np} 
\be \epsilon \equiv
\frac{\mpl^2}{16\pi} \left(\frac{ V'}{V}\right)^2 = \frac {M_{\zt{Pl}}^2 \tanh^2 \lb \phi / M \rb}{16 \pi M^2}\,, \ee 
and 
\bea
\eta  & \equiv &  \frac{\mpl^2} {8\pi} \left| \frac{ V''} {V} - \frac{1}{2}
  \left(\frac{V'}{V} \right)^2\right| \nonumber \\ & = & \frac {M_{\zt{Pl}}^2 |-5 +\cosh \lb 2\phi/M
    \rb| }{32 \pi M^2 \, \cosh^2 \lb \phi /M \rb}\,. \eea
The density fluctuation at scale $k$ is
\bea
\left. \frac{\delta\rho}{\rho}\right|_k &=&\frac{H^2/2\pi}
{(\dot\phi_k)_{\rm f.o.}} \nonumber \\
 &  = &  \frac{(8\pi)^{3/2}} {2\pi\sqrt{3}}\frac{1}{\mpl^3} \left. \left(
    \frac{V^{3/2}}{V'}\right)_k\ \right|_{\rm f.o.}  \\
& = & \left. -\frac {8 M \sqrt{2 \pi/ 3} \coth \lb \phi/ M \rb \sqrt{ V_0/
    \cosh \lb \phi/ M \rb} }{M_{\zt{Pl}}^3} \right|_{\rm f.o.} \,, \nonumber
\label{ocho}
 \eea
where $H = \dot a/a$ is the Hubble parameter and the field $\phi$ is evaluated at freeze out (f.o.), the time when the scale $k$ leaves the horizon.

A straightforward calculation shows that (\ref{V1}) and (\ref{V2}) 
 share similar behavior, namely that $\xe$ and $\eta$ are of the scale
$M_{\zt{Pl}}^2/M^2$. For (\ref{V1})  $\xe$ and $\eta$ quickly grow near
the end of inflation ($\phi \sim M$), whereas for (\ref{V2}) $\xe$ and
$\eta$ remain small. Thus for (\ref{V2}), as for power-law inflation (with an exponential potential), inflation does not end. 
We assume that the dynamics of a second field leads to exit from the inflationary phase into the reheating phase. 

The number of $e$-folds from $t_{\rm i}$ to $t_{\rm e}$ is given by \bea
N({\rm i} \ra {\rm e}) &=& \int_{t_{\rm i}}^{t_{\rm e}}\ H(t) dt
\nonumber \\
& = & -\frac{8\pi}{\mpl^2}\ \int_{\phi_{\rm i}}^{\phi_{\rm e}}\ \frac{V\ d\phi}{V'} \\
& = & \left. \frac {8 M^2 \pi \ln \left[\sinh \lb \frac \phi M
      \rb\right ]} {M_{\zt{Pl}}^2} \right|^{\phi_{\rm e}}_{\phi_{\rm i}} \,
. \nonumber \eea Then, the requirement that there be roughly $50-60$ $e$-folds of
observable inflation yields \be  M \agt 1.4 \, M_{\rm Pl} \, ,
\label{seis}
\ee
where we have neglected logarithmic corrections from the high field contributions to $N$~\cite{Freese:1990rb}.

\begin{table*}
  \caption{
    Model parameters, $n_t$, $dn_s/d\ln k$, and $dn_t/d\ln k$ for $n_s$ and $r$ within their $1\sigma$ uncertainties for the potential of Eq.~(\ref{V1}).
 \label{table}
  }
\begin{tabular}{ccccccccc}
\hline
\hline
~~~~$n_s$~~~~ & ~~~~$r$~~~ & ~~~~$\phi_*/M_{\rm Pl}$~~~~ & ~~~~$M/M_{\rm
  Pl}$~~~~  & ~~~~$V_0/M_{\rm Pl}^4$~~~~ &
~~~~$V(\phi_*)/M_{\rm Pl}^4$~~~~ & ~~~~$n_t$~~~~ &
~~$dn_s/d\ln k$ ~~ & ~~ $dn_t/d\ln k$~~ \\
\hline
 0.95  & 0.20 & $1.24$ & $1.04$ & $1.88\times 10^{-11}$ & $1.04\times 10^{-11}$ & $-0.026$  & 0.00059 & $-0.00058$\\
 0.95 & 0.27 & $1.51$ & $0.982$ & $3.43\times 10^{-11}$ & $1.41\times 10^{-11}$ & $-0.035$  & 0.00048 & $-0.00048$\\
 0.95 & 0.15 & $1.07$ & $1.08$ & $1.19\times 10^{-11}$ & $7.82\times 10^{-12}$ & $-0.019$  & 0.00057 & $-0.00056$\\
 0.96 &  0.20 & $1.58$ & $1.11$ & $2.28\times 10^{-11}$ & $1.04\times 10^{-11}$ & $-0.026$  & 0.00034 & $-0.00034$\\
0.96 & 0.27 & $2.10$ & $1.04$ & $5.35\times 10^{-11}$ & $1.41\times 10^{-11}$ & $-0.034$  & 0.00017 & $-0.00017$\\
0.96 & 0.15 & $1.33$ & $1.17$ & $1.34\times 10^{-11}$ & $7.82\times 10^{-12}$ & $-0.019$  & 0.00038 & $-0.00038$\\
0.97 & 0.20 & $2.35$ & $1.21$ & $3.72\times 10^{-11}$ & $1.04\times 10^{-11}$ & $-0.025$  & 0.00011 & $-0.00011$\\
 0.97 & 0.15 & $1.80$ & $1.28$ & $1.69\times 10^{-11}$ & $7.82\times 10^{-12}$ & $-0.019$  & 0.00020 & $-0.00020$\\
\hline
\hline
\end{tabular}
\end{table*}

The scalar and tensor power spectra are parametrized by \bea {\cal
  P}_\chi & = & A_s \lb \frac k {k_*} \rb^{n_s -1 + \frac 1 2
  \alpha_s  \ln \lb \frac k {k_*} \rb  + \cdots}\, ,   \\
{\cal P}_h & = & A_t \lb \frac k {k_*} \rb^{n_t + \frac 1 2 \alpha_t
  \ln \lb \frac k {k_*} \rb + \cdots}\, . \eea 
Under the approximations of
Ref.~\cite{Leach:2002ar}, the spectral indices and their running are
\begin{eqnarray}
n_s& \simeq & 1-4\epsilon+2\eta+ \left(\frac {10} 3 +4C \right) \xe
\eta - (6+4C) \xe^2 + \frac 2 3 \eta^2 \nn
& & - \frac{2}{3} (3 C-1) \left(2 \xe^2-6 \xe \eta +\xi ^2\right)\,,  \\
n_t &\simeq & -2\epsilon + \left(\frac 8 3 +4C \right) \xe \eta -\frac 2 3 (7+6C) \xe^2 \,,  \\
  \alpha_s & \equiv & {d n_s \over d \ln k}  \simeq  -8 \epsilon ^2+
16 \epsilon \eta -2\xi^2\,,  \\ 
\label{indices}
\alpha_t & \equiv & {d n_t \over d \ln k}  \simeq  -4\epsilon(\epsilon-\eta)\,,
\end{eqnarray}
with  $C = \gamma_{\rm E} +
\ln 2 - 2 \approx -0.7296$. For the potential in \er{V2}, $\xi^2$ is given by 
\bea
\xi^2 & \equiv & \frac {M_{\rm Pl}^4 V' V'''} {64 \pi^2 V^2} \nn
& = & \frac {M_{\rm Pl}^4} {64 \pi^2 M^4} \tanh^2 \lb \frac {\phi_*} M\rb \lsb -5+6 \tanh^2 \lb \frac {\phi_*} M \rb \rsb \label{xi} , \quad\quad
\eea
with $\phi_*$  the field value when the $k_*$ scale crosses the
horizon, $k_* = a H$.
The amplitudes are related to $\xe,\eta$ and $V$ by
\bea
  A_s &\simeq & {8V\over 3M_{\rm{Pl}}^4\epsilon}\left[1 - (4C + 1)
    \epsilon + \lb 2C -\frac 2 3\rb\eta\right]\,, \label{eq:amplitude} \\
A_t & \simeq &   {128V\over 3M_{\rm{Pl}}^4}\left[1 - \left(2C + \frac 5
    3 \right) \epsilon \right]  \, .
\eea
Finally, the ratio of the amplitudes of the spectra at the pivot
$k=k_*$  is
\be
  \label{eq:R}
 r\equiv {A_t \over A_s} \simeq 16 \epsilon + 32\lb C - \frac 1 3 \rb \xe (\xe - \eta) \, .
\ee

\begin{figure}[tb]
\postscript{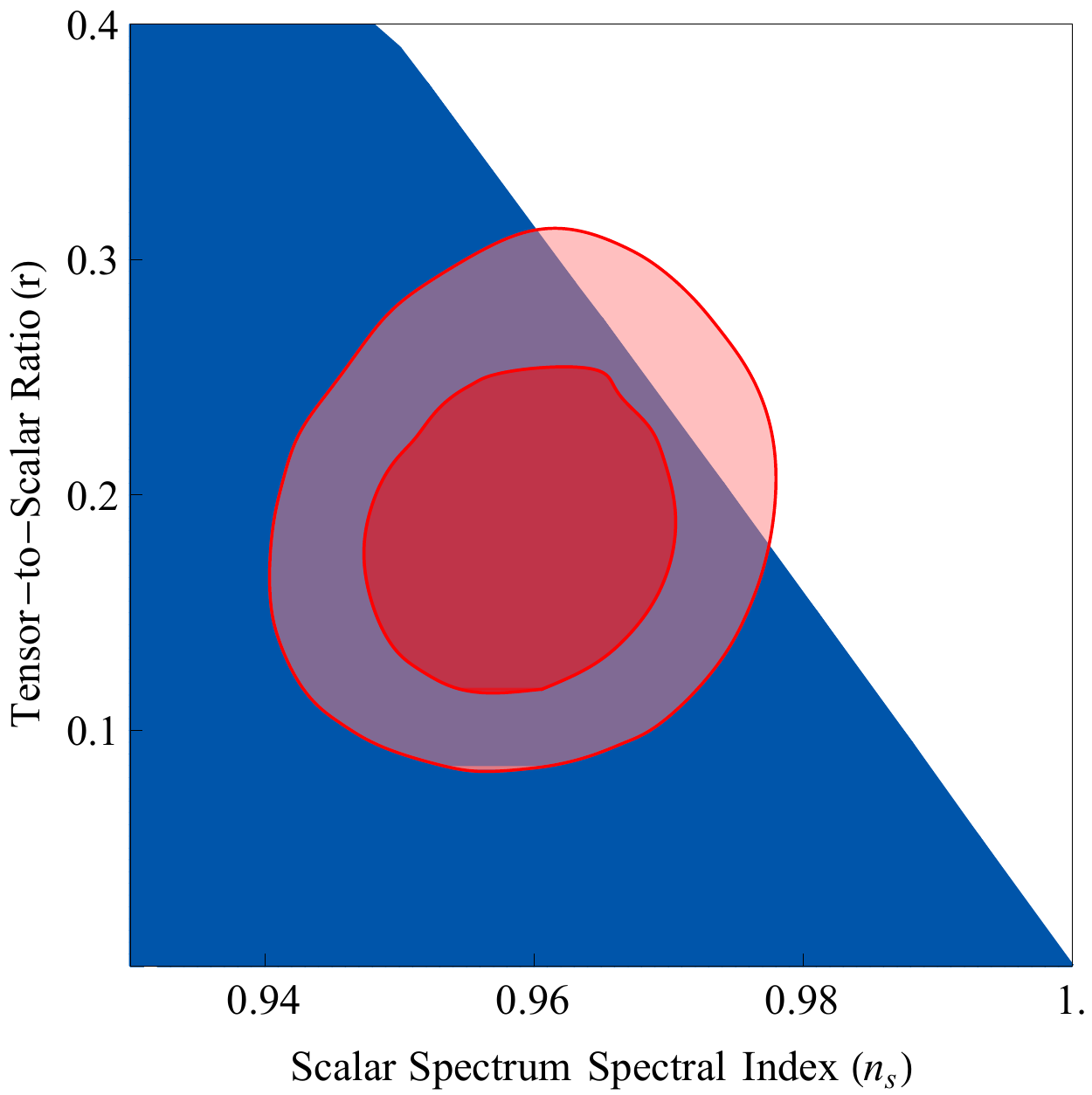}{0.9}
\caption{The blue region shows the parameter space available to the potential of Eq.~(\ref{V1}).
 The red areas shows the 68\%
  and 95\%~C.L. regions favored by a combination of Planck data, WMAP polarization data and small scale CMB data  in combination with the BICEP2 constraint on $r$~\cite{Ade:2014xna}.
 \label{fig1}}
\end{figure} 

The parameters $M$ and $\phi_*$ are determined from the slow-roll
parameters, which in turn are determined by $r$ and $n_s$. $V_0$ is
fixed by the amplitude, $A_s \simeq 22.2 \times
10^{-10}$~\cite{Ade:2013uln}, via Eq.~\er{eq:amplitude}, which gives 
  the density contrast $\delta \rho/\rho \sim 10^{-5}$. The parameter
  space that admits positive solutions for $\phi_*$ and $M$ is shown
  in Fig.~\ref{fig1}.  With $\xe, \eta$ and $\xi$ known, we also
  compute $n_t$, and the running of the spectral indices.  In
  Table~\ref{table}, we display the parameters of the model together
  with $n_t$, $\alpha_s$ and $\alpha_t$ for values of $n_s$ and $r$
  within their $1\sigma$ uncertainties. The values of $M$ agree with
  (\ref{seis}) modulo logarithmic corrections from the high field
  contributions to $N$. The predictions for $\alpha_s$  are within the 95\% C.L. region favored by Planck data~\cite{Ade:2013uln}. Note that $[V(\phi_*)]^{1/4}$ is of the order of $10^{16}$ GeV, the grand unification scale.

\begin{table*}
  \caption{
    Model parameters, $n_t$, $dn_s/d\ln k$, and $dn_t/d\ln k$ for
    $n_s$ and $r$ within their $1\sigma$ uncertainties for the potential of Eq.~(\ref{V3}). \label{table2}
  }
\begin{tabular}{ccccccccc}
\hline
\hline
~~~~$n_s$~~~~ & ~~~~$r$~~~ & ~~~~$\phi_*/M_{\rm Pl}$~~~~ & ~~~~$M/M_{\rm
  Pl}$~~~~  & ~~~~$V_0/M_{\rm Pl}^4$~~~~ &
~~~~$V(\phi_*)/M_{\rm Pl}^4$~~~~ & ~~~~$n_t$~~~~ &
~~$dn_s/d\ln k$ ~~ & ~~ $dn_t/d\ln k$~~ \\
\hline
$0.958$ & $0.19$ & $2.64$ & $5.40$ & $1.09 \times 10^{-11}$ & $ 9.72
\times 10^{-12}$ & $-0.024$ &
$-0.00056$ & $-0.00043$ \\
$0.950$ & $0.15$ & $1.32$ & $4.32$ & $7.92 \times 10^{-12}$ & $7.56
\times  10^{-12}$ &
$-0.019$ & $-0.00024$ & $-0.00058$\\
$0.960$ & $0.15$ & $1.87$ & $4.99$ & $8.18 \times 10^{-12}$ & $7.64 \times 10^{-12}$ &
$-0.019$ & $-0.00028$ & $-0.00040$\\
$0.970$ & $0.15$ & $4.66$ & $2.26$ & $3.07 \times 10^{-11}$  & $7.73
\times 10^{-12}$ & $-0.019$ & $+0.00008$ & $-0.00021$\\
\hline
\hline
\end{tabular}
\end{table*}

\begin{figure}[t]
\postscript{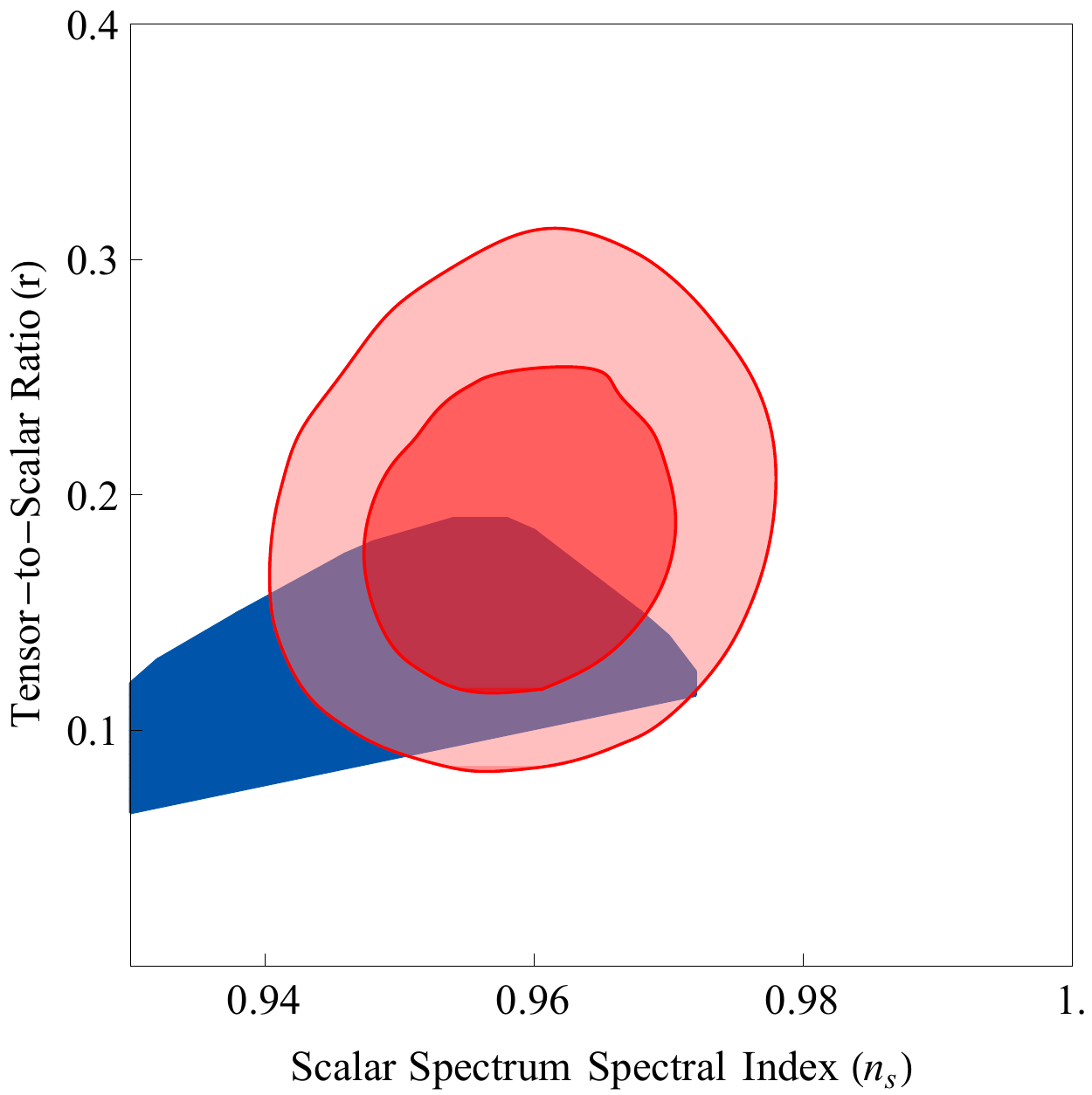}{0.9}
\caption{Similar to Fig.~\ref{fig1} for the potential of Eq.~(\ref{V3}) with $50-60$ e-folds of inflation.
 \label{fig2}}
\end{figure} 

An $S$-dual potential for which inflation ends (and that solves the unlikeliness problem) is
\begin{equation}
V=V_0\ \lsb {\rm sech}(3\phi/M) - \frac 1 4 {\rm sech}^2(\phi/M)\rsb
\, .
\label{V3}
\end{equation}
The number of $e$-folds from the time of horizon crossing $t_*$ to the end of inflation $t_{\rm e}$ when $\xe(\phi_{\rm e})=1$ is given by \bea N_* &=&  -\frac {8\pi M^2} {M_{\zt{Pl}}^2} \int^{y_{\rm e}}_{y_*} dy \\
  & \times & \frac{\left(3 y^2+1\right) \left(3 y^2-4
      \sqrt{1-y^2}+1\right)}{4 y \left(y^2-1\right) \left(\frac{1}{2}
      \left(3 y^2+1\right)^2-3 \sqrt{1-y^2} \left(y^2+3\right)\right)}
  \, , \nonumber \eea where
   $y \equiv \tanh (\phi/ M)$. Inflation ends before $V$ is rendered negative by the
  $-\zt{sech}^2 (\phi/M)/4$ term. In Fig.~\ref{fig2} and Table~\ref{table2}, we display the
allowed region and various parameters for a few allowed values of $n_s$ and $r$.

{\it Acknowledgements.}  This work was supported by the DOE under
Grant Nos. DE-FG02-13ER42024 and DE-FG02-95ER40896, by the NSF under
Grant Nos. PHY-0757959 and PHY-1053663, and by NASA under Grant
No. NNX13AH52G.



\begin{thebibliography}{99}

\bibitem{Ade:2014xna} 
  P.~A.~R.~Ade {\it et al.}  [BICEP2 Collaboration],
  arXiv:1403.3985 [astro-ph.CO].

\bibitem{Hinshaw:2012aka} 
  G.~Hinshaw {\it et al.}  [WMAP Collaboration],
  Astrophys.\ J.\ Suppl.\  {\bf 208}, 19 (2013)
  [arXiv:1212.5226 [astro-ph.CO]].

\bibitem{Sievers:2013ica} 
  J.~L.~Sievers {\it et al.}  [Atacama Cosmology Telescope Collaboration],
  JCAP {\bf 1310}, 060 (2013)
  [arXiv:1301.0824 [astro-ph.CO]].



\bibitem{Keisler:2011aw} 
  R.~Keisler, C.~L.~Reichardt, K.~A.~Aird, B.~A.~Benson, L.~E.~Bleem, J.~E.~Carlstrom, C.~L.~Chang and H.~M.~Cho {\it et al.},
  Astrophys.\ J.\  {\bf 743}, 28 (2011)
  [arXiv:1105.3182 [astro-ph.CO]].



\bibitem{Ade:2013uln} 
  P.~A.~R.~Ade {\it et al.}  [Planck Collaboration],
  arXiv:1303.5082 [astro-ph.CO].





\bibitem{Ijjas:2013vea} 
  A.~Ijjas, P.~J.~Steinhardt and A.~Loeb,
  Phys.\ Lett.\ B {\bf 723}, 261 (2013)
  [arXiv:1304.2785 [astro-ph.CO]].


\bibitem{Liu:2014mpa} 
  H.~Liu, P.~Mertsch and S.~Sarkar,
  arXiv:1404.1899 [astro-ph.CO].

\bibitem{Montonen:1977sn} 
  C.~Montonen and D.~I.~Olive,
  Phys.\ Lett.\ B {\bf 72}, 117 (1977).

 
\bibitem{Font:1990gx}  See, {\em e.g.}
  A.~Font, L.~E.~Iba\~nez, D.~L\"ust and F.~Quevedo,
  Phys.\ Lett.\ B {\bf 249}, 35 (1990);
  A.~Sen,
  Int.\ J.\ Mod.\ Phys.\ A {\bf 9}, 3707 (1994)
  [hep-th/9402002];
  K.~Becker, M.~Becker and J.~H.~Schwarz,
  {\it String theory and M-theory: A modern introduction,}
  (Cambridge, UK: Cambridge Univ. Pr., 2007).


\bibitem{Lidsey:1995np} See {\it e.g.},
  J.~E.~Lidsey, A.~R.~Liddle, E.~W.~Kolb, E.~J.~Copeland, T.~Barreiro and M.~Abney,
  Rev.\ Mod.\ Phys.\  {\bf 69}, 373 (1997)
  [astro-ph/9508078];
  D.~Baumann,
  arXiv:0907.5424 [hep-th].


\bibitem{Freese:1990rb} We note that inflationary potentials can be characterized by a scale $M > M_{\rm Pl}$ provided the energy
  \mbox{density~$\ll M_{\rm Pl}^4$}. A well-known example is natural inflation where the axion scale $f > M_{\rm Pl}$, see K.~Freese, J.~A.~Frieman and A.~V.~Olinto,
  Phys.\ Rev.\ Lett.\  {\bf 65}, 3233 (1990).
  For a string/supergravity example, where the
  choice of scale is dictated by the compactification topology, see
  S.~Kachru, R.~Kallosh, A.~D.~Linde and S.~P.~Trivedi,
  Phys.\ Rev.\ D {\bf 68}, 046005 (2003)
  [hep-th/0301240].
In their illustrated example the chocice of the parameter
  $a =0.1$ corresponds to $M \simeq 2 M_{\rm Pl}$.



\bibitem{Leach:2002ar} 
  S.~M.~Leach, A.~R.~Liddle, J.~Martin and D.~J Schwarz,
  Phys.\ Rev.\ D {\bf 66}, 023515 (2002)
  [astro-ph/0202094].

\end{thebibliography}
\end{document}